\documentclass[twocolumn]{aastex62}

\newcommand{\gray}{$\gamma$-ray}
\newcommand{\grays}{$\gamma$-rays}
\newcommand{\bl}{TXS 0506+056}
\usepackage{amsmath}

\received{-}
\revised{-}
\accepted{-}
\submitjournal{ApJ}

\shorttitle{$\gamma$-rays and neutrinos from TXS 0506+056}
\shortauthors{Sahakyan}

\begin{document}

\title{Lepto-hadronic $\gamma$-ray and neutrino emission from the jet of TXS 0506+056}

\correspondingauthor{Narek Sahakyan}
\email{narek@icra.it}
\author[0000-0003-2011-2731]{Narek Sahakyan}
\affiliation{ICRANet-Armenia, Marshall Baghramian Avenue 24a, Yerevan 0019, Armenia.}
\affiliation{ICRANet, P.zza della Repubblica 10, 65122 Pescara, Italy.}



\begin{abstract}
The observation of IceCube-170922A event from the direction of TXS 0506+056 when it was in its enhanced $\gamma$-ray emission state offers a unique opportunity to investigate the lepto-hadronic processes in blazar jets. Here, the observed broadband emission of TXS 0506+056 is explained by boosted synchrotron/synchrotron self Compton emission from the jet whereas the $\gamma$-ray data observed during the neutrino emission- by inelastic interactions of the jet-accelerated protons in a dense gaseous target. The proton energy distribution is $\sim E_{p}^{-2.50}$, calculated straightforwardly from the data obtained by Fermi-LAT and MAGIC and if such distribution continues up to $E_{c,p}=10$ PeV, the expected neutrino rate is as high as $\sim0.46$ events during the long active phase of the source or $\sim0.15$ if the activity lasts 60 days. In this interpretation, the energy content of the protons above $>$ GeV in blazar jets can be estimated as well: the required proton injection luminosity is $\simeq2.0\times10^{48}\:{\rm erg\:s^{-1}}$ exceeding $10^3$ times that of electrons $\simeq10^{45}\:{\rm erg\:s^{-1}}$ which are in equipartition with the magnetic field. As the required parameters are physically realistic, this can be an acceptable model for explanation of the neutrino and $\gamma$-ray emission from TXS 0506+056.
\end{abstract}



\section{Introduction}\label{sec:1}
Recently, IceCube detected Very High Energy (VHE; $>100$ GeV) neutrinos from extragalactic sources \citep{neutrino1,neutrino2}, \citep{neutrino3}. This opens new perspectives for investigation of nonthermal processes in astrophysical objects even though no sources emitting these neutrinos have been identified so far. As neutrinos are not absorbed when interacting either with a photon field or the matter, unlike the \grays{}, they can be detected from distant sources which are "Terra Incognita" for \gray{} observations.\\
Different types of objects are proposed as potential sources for VHE neutrino emission among which the most prominent are Active Galactic Nuclei (AGNs). AGNs, being powered by supermassive black holes, are among the most luminous and energetic extragalactic objects. When the jet of an AGN is aligned with an observer's line of sight, they appear as blazars \citep{urry} which, based on the observed emission lines, are commonly sub-divided into BL Lacertae (BL-Lac) objects and Flat-Spectrum Radio Quasars (FSRQs). Blazars are characterized by high luminosity (e.g., $10^{48}-10^{49}\:{\rm erg\:s^{-1}}$ in the \gray{} band) and variable nonthermal emission in the radio to the VHE \gray{} bands. As blazar emission is dominating in the extragalactic \gray{} sky, it is natural to consider them as the most probable sources of the observed neutrinos. In fact, \citet{padovani1, padovani2}, have recently found a correlation between High Energy (HE) peaked BL Lacs (emitting $> 50$ GeV) detected by Fermi Large Area Telescope (Fermi-LAT) and the VHE neutrino sample detected by IceCube. Also, remarkable is the possible association of the highest-reconstructed-energy ($\sim2$ PeV) neutrino event with the exceptionally bright phase of FSRQ PKS B1414-418 \citep{kadler}. Recently, \citet{aartsenB} showed that the maximum contribution of the known blazars to the observed astrophysical neutrino flux in the energy range between 10 TeV and 2 PeV is less than 27 \%, but in principle significant neutrino emission from a particularly bright blazar can be detected.\\
There are proposed different mechanisms for VHE neutrino production in blazar jets (e.g., \citet{man, bednarekneutrino, prot, yoshida, tavec}, \citep{ 2016PhRvL.116g1101M, 2016PhRvD..93h3005W, 2016MNRAS.455..838K}). These relativistic jets are ideal laboratories where the leptons (electrons) and hadrons (protons) can be effectively accelerated, which interacting with the magnetic and/or photon fields can produce emission across the whole electromagnetic spectrum. Even if the leptonic emission alone can explain the observed features of some blazars, the energetic protons co-accelerated with electrons might contribute to the observed emission. As the protons probably carry a significant fraction of the total jet power, the exact estimation of their content in the jet is crucial for understanding the jet launching, collimation and dynamics. In general, only HE \gray{} observations alone are not sufficient to differentiate between the contribution of protons and electrons; this can be done only by neutrino observations.\\
The lack of a high-confidence association of a neutrino event with a particular blazar significantly complicates the interpretation of hadronic emission from blazar jets. The best association to date is that between IceCube-170922A neutrino event with the \gray{} bright BL Lac object \bl{} \citep{IceCubeFermi}. The Fermi LAT and MAGIC observations reveal that \bl{} was in the active state in MeV/GeV and above 100 GeV bands when the neutrino event was observed on September 22, 2017; the evolution of its multiwavelegth emission in time around the neutrino event can be seen here \href{https://youtu.be/lFBciGIT0mE}{\nolinkurl{youtu.be/lFBciGIT0mE}} \citep{sah}. At the redshift of $z=0.336$ \citep{paiano} \bl{} is among the brightest BL Lacs detected by Fermi LAT. Moreover, IceCube detected a $3.5\sigma$ excess of neutrinos from the direction of \bl{} in 2014-2015 \citep{IceCube1}. Further detailed spatial and temporal analyses of the complex \gray{} region around \bl{} showed that the emission from the nearby flaring blazar PKS 0502+049 is dominating at low energies, but \bl{} is brighter above a few GeV, making it the most probable neutrino source \citep{sah}.\\
The data available from the observations of \bl{} makes it a unique object for testing the lepto-hadronic emission scenarios in blazar jets. For example, in \citet{magic} the one-zone lepto-hadronic model based on the interaction of both accelerated electrons and protons (photo-meson reaction) with the external photons (from a slow-moving external layer) can successfully explain the observed multiwavelength SED and neutrino rate if the proton energies are in the range from $10^{14}$ eV to $10^{18}$ eV. In this and other similar models it is impossible to estimate the relative contribution of low-energy protons ($>1$ GeV) which carry a significant portion of the jet power. An alternative modeling of the multiwavelength emission from \bl{} is applied in the current paper; it is assumed that the protons accelerated in the jet of \bl{} interact with a dense target crossing the jet and produce the observed HE and VHE \gray{} emission from the decay of neutral pions. This is done within widely discussed jet-target interaction models which were successfully applied for modeling the emission from different AGNs (e.g., M87 \citep{barkov, 2012ApJ...755..170B}, Cen A core \citep{araudo10}, Mrk 421 \citep{dar}, 3C454.3 \citep{2013ApJ...774..113K},  etc.). The initial proton distribution is estimated by normalizing the expected \gray{} flux from proton-proton ($pp$) interactions to the Fermi LAT and MAGIC data, then the neutrino spectra are estimated straightforwardly. There is no contradiction between this and other discussed models involving photo-meson reactions as, again, the cascade initiated from the interaction of ultrahigh-energy protons might be still responsible for the emission in the X-ray band. By interpreting that the observed HE and VHE \gray{} emission originates from $pp$ interactions, the main purpose of the current paper is to estimate {\it i)} the total luminosity of protons ($>$ GeV) and compare it with that of electrons and {\it ii)} the expected detection rate of neutrinos produced from $pp$ inelastic collisions.\\
The paper is structured as follows: the model adopted here is described in Section \ref{sec:2}. The modeling of broadband emission from \bl{} is presented in Section \ref{sec:3}. The results are discussed in Section \ref{sec:4} and summarized in Section \ref{sec:5}.
\section{Broadband emission from jets}\label{sec:2}
Dominantly, the emission from blazars is of a non-thermal nature, extending from radio to VHE \gray{} bands, with two broad humps in the spectral energy distribution (SED), peaking in the IR-X-ray and MeV-TeV bands. This double-peaked feature can be attributed to radiative losses of non-thermal electrons. The low-energy component can be well explained by synchrotron emission of relativistic electrons in the jet and the inverse Compton scattering of soft target photons on these electrons can be responsible for the HE peak. The target photon fields depend mostly on the location of the emission region \citep{sikora09}, being of different origin and varying along the jet. For example, for BL Lacs with very weak (or absent) emission lines, the target photons can be synchrotron photons (synchrotron self Compton (SSC) \citep{ghisellini, maraschi, bloom}), whereas for FSRQs where the jet propagates in an environment reach of photons, the HE component can originate also from inverse Compton scattering of external photons (e.g., photons reflected from Broad Line Region (BLR) \citep{sikora} or from a dusty torus \citep{blazejowski,ghiselini09}). These models have been successful in explaining several features observed in the non-thermal spectra of blazars, e.g., spectra in different bands, simultaneous or non-simultaneous flux increases (decreases), etc. However, the observation of minute-scale \gray{} variability of blazars poses severe difficulties for one-zone leptonic scenarios; the emission region will be beyond the event horizon of the central black holes only in case of extremely high bulk Lorentz factors (e.g., see \cite{2012ApJ...749..119B,2017ApJ...841...61A}). Therefore, there have been proposed more complex scenarios: multi-zone models (e.g., spine-sheath \citep{tavecchio08}, decelerating-jet \citep{georganopoulos}, jets in a jet \citep{giannios09,giannios10}, etc.) and internal-shock models (e.g., \citep{marscher,spada,sokolov,joshi,dermer10}).\\
The HE component in blazar SEDs can be also explained by applying alternative models invoking the radiative output of hadrons accelerated in the jets of blazars. The protons unavoidably accelerated in jets can interact with matter (proton-proton [$pp$]), magnetic or radiation field (proton-$\gamma$) and produce the observed HE emission. In the last two cases, in order to emit either through proton-synchrotron emission \citep{ahartev,muckesynch} or photo-pion production \citep{mannheim}, the protons should be accelerated to extremely high energies ($E_{p}\geq10^{19}$eV) and propagate in a highly magnetized plasma ($B\geq30$ G). These requirements for the protons and/or the medium can be somewhat softened when a highly collimated proton beam accelerated in the jet penetrates into a dense and compact target (e. g., cloud(s) from BLR \citet{dar,beall,araudo10}) and produces the observed HE \grays{} through inelastic $pp$ scattering. In addition, at larger distances from the nucleus rich in star populations, the star-jet interactions too can produce HE \grays{} \citep{barkov,araudo13, beall, bednarek97, bednarek15, cita}. In these cases, the protons should be accelerated only up to moderate $\simeq100$ TeV energies unlike the other models.
\subsection{The model}
The interaction of the blazar jets with clouds or stars abundant in their environment does not significantly affect the dynamics of the jet and its propagation to $kpc$ scales, but it provides targets for efficient hadronic interactions. The strong shock formed at interaction of relativistic flows with dense targets can accelerate the particles and in some cases their interactions can be responsible for the steady (e.g., several clouds can interact with the jet simultaneously) or flare-like \gray{} emissions. The studies of hydrodynamical simulations of jet-target interactions and dynamical time scales required for particle acceleration and emission show that star-target interactions can produce detectable \gray{} fluxes not only in the nearby (e.g., Cen A \citep{araudo10} and  M87 \citep{barkov}) but also in distant/powerful objects (e.g., 3C 454 \citep{2013ApJ...774..113K}, Mrk 421 \citep{beall,dar}). For a comprehensive and detailed study of the hydrodynamical simulation of jet-cloud/star interactions and discussion of jet stability see \citet{2012A&A...539A..69B} and \citet{2017A&A...606A..40P}.\\
The scenarios mentioned above require several parameters for accurate estimation of the duration, rate and efficiency of interactions and the related radiative outputs. The parameters describing the targets (e.g., clouds or stars envelopes) and the energy distribution of accelerated protons are different for each source, but even within the typical values (e.g., clouds with $\sim10^{10}-10^{12}\:{cm^{-3}}$ density and $\sim10^{12}-10^{13}\:{cm}$ radius) the powerful jet-target interactions can produce observable \gray{} emission. Another important parameter is the distance ($d$) from the base of the jet where the interaction occurs. Star-jet/cloud interaction cannot occur within the jet formation region and also at large distances from the central source, as the jet energy flux decreases with the distance as $F_{j}\simeq L_j/\pi \theta^2 d^2 $ ($\theta$ is the jet opening angle). For very powerful jets the jet energy flux might be still high enough for an effective interaction farther from the jet launching point which will also allow the protons to be accelerated to HEs. On the other hand, for low-power jets, when the penetration occurs in the innermost part of the jet, the protons can be further accelerated in the target, gaining energies required for HE \gray{} emission \citep{araudo10,barkov}. Even though the jet-target interactions can occur frequently, for some jets the observed \gray{} emission is likely to be dominated by boosted SSC emission from jet-accelerated electrons. However, in some cases when the leptons cool faster not reaching HEs or due to some internal changes in the jet affecting only the boosted emission, the hadronic component produced in star-jet interactions can dominate and produce the observed HE and VHE \grays{}. Such scenario can be applied only when \gray{} and neutrino emissions are detected, e.g., from \bl.\\
Since our primary goal is to investigate whether the observed HE and VHE \gray{} and neutrino emission from \bl{} can be explained by the interaction of moderately accelerated protons and measure their content in the jet, here, without going much in details (e.g., the nature of target, at which distance the interaction occurred, etc.), the broadband SED of \bl{} obtained during the neutrino event detection is modeled. Assuming that the energetic protons (either accelerated in the jet or in the target) interact in a target with a typical density of $10^{10}{cm^{-3}}$, the \grays{} are produced through $pp$ interactions. Most likely, the maximum energy of protons will be defined by the size of acceleration region rather than by radiative losses, i.e. at a rate close to the theoretical limit $t_{\rm acc} \sim R/c$. Expressing the acceleration time as $E/dE/dt$, where $dE/dt\simeq \eta e B c$ \citep{2002PhRvD..66b3005A} is the proton acceleration rate with $\eta$ efficiency, implies that $E_{\rm max}\simeq 3.0\times10^{15}\:(\eta/0.1)\:(B/1{\rm G})\:(R/10^{13}{\rm cm})$ eV. So, even if the proton acceleration occurs in a relatively small ($\sim10^{13}$ cm) dense target, $E_{\rm}$ goes well beyond $\sim10$ PeV, enough to produce the observed VHE \gray{} photons and neutrino. In the case when the protons are accelerated in the jet, in the frame of the target their energy will be even higher due to Doppler boosting. The protons interacting in the target produce \grays{} from the decay of neutral pions ($\pi^0\rightarrow \gamma \gamma$), while the neutrinos ($\nu_\mu, \nu_e$) are produced from the decay of $\pi^\pm$ (e.g., $\pi^+ \rightarrow \mu^+ + \nu_\mu \rightarrow e^+ + \nu_e + \nu_\mu +{\bar \nu}_\mu$). The observed \gray{} luminosity of \bl{} around 10 GeV (most likely the peak of HE component) is $\simeq4\times10^{46}\:{\rm erg\:s^{-1}}$, then assuming the efficiency of energy transfer from relativistic protons to secondary particles is $\sim10\%$, the required proton luminosity should be $L_{\rm p}\simeq4\times10^{47}\:{\rm erg\:s^{-1}}$. Next, assuming the proton acceleration efficiency is roughly 10\%, the jet power would be $L_{\rm jet}\simeq4\times10^{48}\:{\rm erg\:s^{-1}}$- a value usually estimated for bright blazars. Thus, both the maximum energy of protons and the required luminosity are physically realistic and the observed HE and VHE \gray{} emission from \bl{} can be due to $pp$ interactions.
\section{SED modeling}\label{sec:3}
The broadband SEDs of \bl{} for low (period 1 [P1]) and high (period 2 [P2]) VHE \gray{} emitting states ({when the neutrino was detected) are shown in Fig. \ref{sed} (taken from \citep{magic}). As compared with the archival data (light gray), {\it i)} the low energy component is relatively constant and {\it ii)} the HE component slightly increases and the spectrum extends above several hundreds of GeV. This well fits in the scenario discussed above; the synchrotron emission of electrons accelerated in the jet is almost unchanged while most likely a different process is responsible for the HE emission in the active states. It is natural also to expect that the \gray{} flux produced by the same electrons remains the same as the time-averaged one.\\
The emissions from two different jet regions are considered in the SED modeling: {\it i)} the observed low-energy and time-averaged HE components are explained as emission directly from jet-accelerated electrons in a compact region moving with the bulk Lorentz factor of the jet and {\it ii)} during the active \gray{} emitting state, when the neutrino was observed, the \grays{} are produced from the inelastic $pp$ interactions in the second emission site. Such a compact emitting region is usually considered when modeling the multiwavelength emission from blazars (e.g., \citep{ghisellini, maraschi, bloom}), while the second zone is expected to form when a target penetrates into the jet (see the sketch in Fig. 1 of \cite{2012A&A...539A..69B} where a jet-target interaction is illustrated). Next, the emission from the $e^{-}e^{+}$ pairs produced from the decay of charged pions in the target, which can be significant in the X-ray band, is also taken into account. This modeling allows to estimate both the electron and proton content in the jet.\\
The radiative contribution of electrons is computed within the one-zone scenario, assuming the emission region (the "blob") is a sphere with a radius of $R$ moving with a bulk Lorentz factor of $\Gamma$, carrying a magnetic field of $B$ and having a population of relativistic electrons with an energy distribution of
\begin{equation}
N^{\prime}_{\rm e}(E^{\prime}_{\rm e})\thicksim E^{\prime \alpha}_{e}\:\exp(-\frac{E^{\prime}_{\rm e}}{E^{\prime}_{\rm cut}})
\label{PLEX}
\end{equation}
between $E^{\prime}_{\rm min}$ and $E^{\prime}_{\rm max}$ \citep{inou}; $E^{\prime}_{\rm cut}$ is the cut-off energy. As the contemporaneous data from the rising  part of the low energy component are missing, the electron spectral index $\alpha$ can not be measured, so $\alpha=2$ expected from acceleration theories is used. Also, $E_{\rm min}^{\prime}=100\:{\rm MeV}$ is used, so the model does not overpredict the archival radio data. The luminosity of the synchrotron emission of these electrons will be amplified by a relativistic Doppler factor of $\delta$ ($\delta=\Gamma$ for small jet viewing angles), for which $\delta=25$ was used, a characteristic value for bright blazars \citep{ghistav}. The observations of \bl{} in the X-ray, HE and VHE \gray{} bands allowed to infer a variability time scale of $t_{\rm d}\leq10^{5}$ s (e.g., \citep{2018arXiv180704537K,magic} which implies that the emission is produced in sub-parsec regions of the jet, so $R\simeq7.5\times10^{16}$ cm will be used.\\
The energy distribution of protons accelerated in the jet can be also expressed as
\begin{equation}
N_{p}(E_p) \thicksim E_{p}^{-\alpha_{p}} \exp\left(-\frac{E_{p}}{E_{c,p}}\right)
\label{fin}
\end{equation}
where $E_{c,p}$ is first considered as a free parameter then $E_{c,p}=10$ PeV is used as the \gray{} data there is no sign of any spectral cutoff. \grays{}, neutrinos and $e^{-}e^{+}$ pairs will be produced in the interactions of these protons. The characteristic cooling time of $pp$ collisions is $t_{pp}\simeq(K \sigma_{pp} n_{H})^{-1}\simeq10^{15}/n_{H}$ which is inversely proportional to the target particle number density. In case of optimal radiation $t_{\rm pp}\simeq t_{\rm v}$ ($t_{\rm v}=10^{5}$ s) the target density is $n_{\rm H}=10^{10}\:{\rm cm^{-3}}$. The size of the emitting target can be estimated as $r_{\rm}=(3 M_{\rm c}\:t_{\rm v}/4\:\pi 10^{15}\:m_{p})^{1/3}\simeq5.2\times10^{13}\:(M_{c}/10^{28} g)^{1/3}\:{\rm cm}$ (e.g., from \citet{barkov} for the weak tidal disruption). We note that this number density and size are typically estimated for the clouds in BLR (e.g., \citep{2015ARA&A..53..365N, 2006LNP...693...77P}) and adopted in star-cloud (e.g., \citet{araudo10}) or star-jet (e.g. \citet{barkov}) scenarios. As the target density is high, the protons lose a significant fraction of their energy at $pp$ collisions: the interaction is in a radiatively efficient regime, $t_{pp}\leq t_{\rm v}$, so most of the \grays{} are emitted around $t_{\rm v}$ rather than when the target is already accelerated to high velocities. The \gray{} and neutrino spectra above 100 GeV are calculated using the analytic approximations from \citet{kelner_06}, while the delta function approximation for lower energies is used (for exact formula see \citet{sahakyan}). The secondary $e^{-}e^{+}$ pairs spectrum will be similar to that of \grays{} but shifted to lower energies (see Eq. 62 in \citet{kelner_06}).\\
During the fitting (synchrotron/SSC emission of jet-accelerated electrons and $pp$ interactions) the model free parameters are derived by Markov Monte Carlo Chain sampling of their likelihood distributions \citep{zabalza}. This allows to obtain the parameters and their uncertainties which best describe the observed spectra statistically. The following expected ranges are considered: $1.5\leq\alpha_{\rm p}\leq10$, $0.511\:{\rm MeV}\leq E^\prime_{cut}, E_{c,p}\leq10\:{\rm TeV}$, and the normalization of electrons/protons and $B$ are defined as positive parameters.
\begin{figure*}
  \centering
    \includegraphics[width= 0.475 \textwidth]{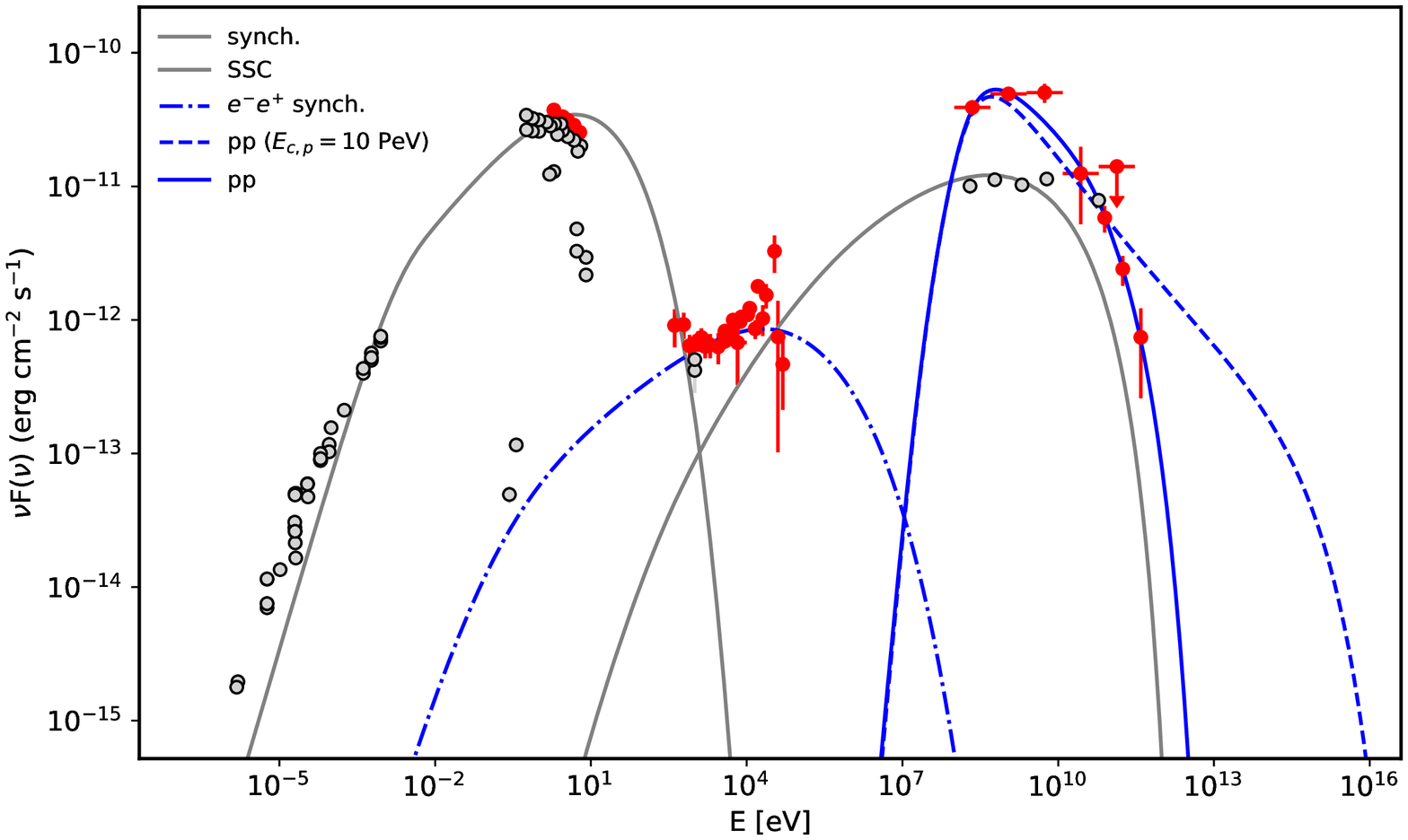}
    \includegraphics[width= 0.475 \textwidth]{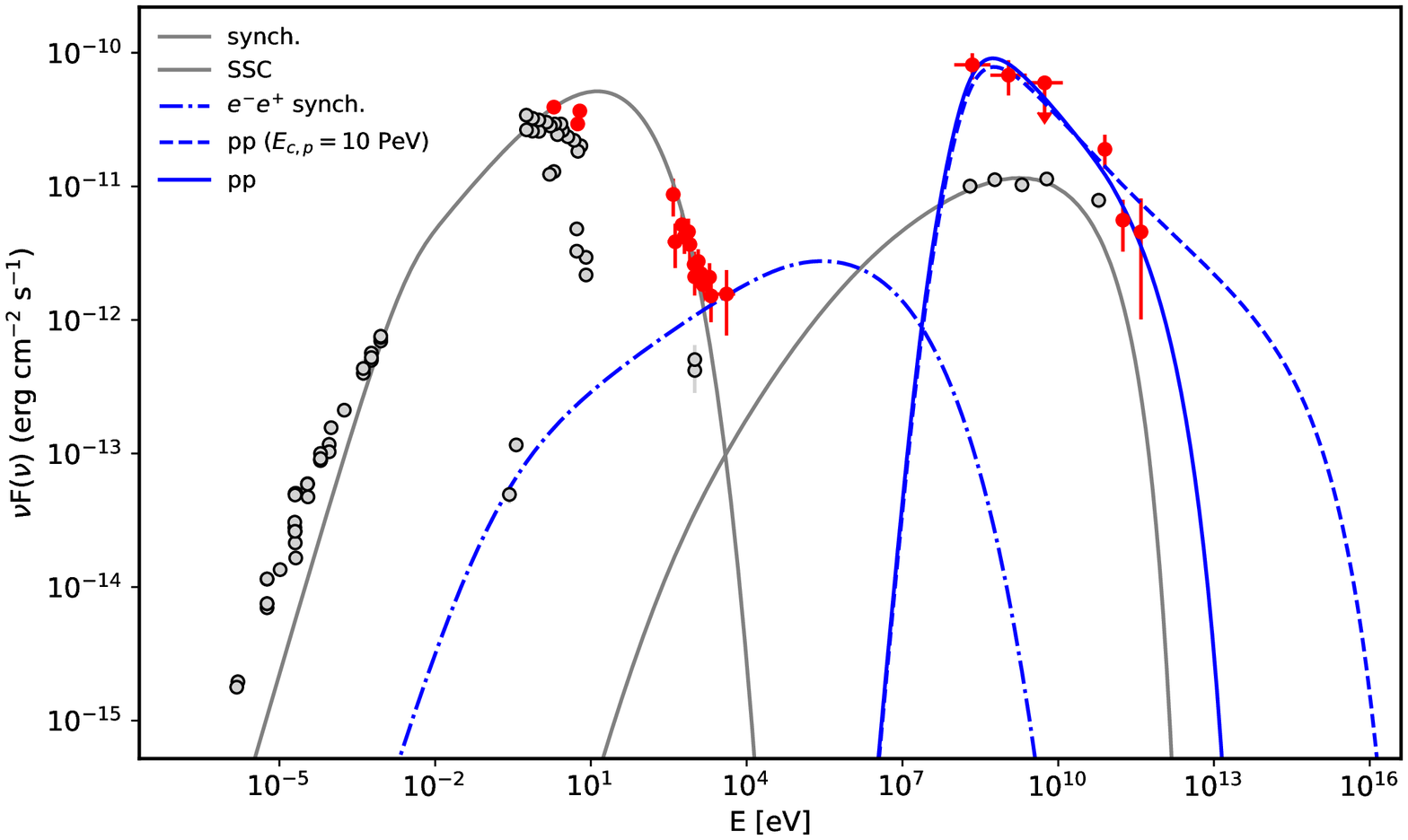}
    \caption{The broadband SED of \bl{} for the active (left) and low VHE \gray{} emitting states. The boosted synchrotron/SSC emission from the jet are shown in gray whereas the \gray{} spectra from {\it pp} interactions and synchrotron emission from secondary $e^{-}e^{+}$ pairs are shown with blue color. The data are corrected for EBL absorption adopting the model from \citet{dominguez}.}%
    \label{sed}
\end{figure*}
\section{Results and Discussions}\label{sec:4}
Detailed temporal and spatial analyses of the complex \gray{} region around the VHE neutrino event IceCube-170922A illustrates that most likely \bl{} is the source of the observed VHE neutrinos \citep{sah}. This is a direct evidence that cosmic rays (protons) are effectively accelerated in the jet of \bl{}. The data available from the multiwavelength observations of \bl{} around the neutrino event make this source an interesting (unique) object where the HE and VHE processes can be investigated using \grays{} as well as neutrinos.\\
The multiwavelegth SEDs modeled within the combined leptonic/hadronic scenario are shown in Fig. \ref{sed}. The boosted synchrotron emission from the jet-accelerated electrons can explain the low-energy component, while the SSC radiation of the same electrons- the time averaged \gray{} data (solid gray line in Fig. \ref{sed}). During P2, the X-ray photon index measured by Swift XRT is relatively steep ($\Gamma_{X}>2.5$), corresponding to the HE tail of synchrotron emission (gray line in the right panel of Fig. \ref{sed}) allowing to estimate the cutoff energy at $E_{\rm c}=8.90\pm0.22$ GeV. The magnetic field energy density is $U_{\rm B}=0.88\times10^{-3}{\rm erg\;cm^{-3}}$ (for $B=0.15\pm0.004$ G) being of the same order as that of the electrons, $U_{\rm e}=1.37\times10^{-3}{\rm erg\;cm^{-3}}$, meaning the system is close to equipartition. During P1, the X-ray flux that most like is produced by a different component, limits the cutoff energy at $6.65\pm0.21$ GeV and magnetic field to $B=0.10\pm0.002$ G, otherwise the model predicted flux will exceed the observed data.  The total jet power in the form of magnetic field and electron kinetic energy calculated as $L_{\rm jet}=\pi c R_b^2 \Gamma^2 (U_{B}+U_{e})$ changes in the range of $L_{\rm jet}=(7.47-9.89)\times10^{44}\:{\rm erg\:s^{-1}}$.\\
The \grays{} produced from the decay of $\pi^{0}$ is shown with blue solid and dashed lines in Fig. \ref{sed}. In this case, the the slope of proton distribution, $\alpha_{\rm p}=2.41\pm0.11$ and $\alpha_{\rm p}=2.49\pm0.07$ for P1 and P2 , respectively, is directly measured from the \gray{} spectra. As the slopes are relatively steep ($\alpha_{p}\simeq2.5$) most of the proton energy output is at lower energies, so the \gray{} emission in the GeV band is mostly dominated by the decay of $\pi^{0}$ with no or a negligible contribution from $e^{-}e^{+}$ pairs from $\pi^{\pm}$ decay or $\gamma\gamma$ absorption. When the cut-off energy in Eq. \ref{fin} is taken as a free parameter, it is constrained by the last point in the \gray{} data (with large statistical uncertainty) and corresponds to $E_{c,p}=0.98_{-0.37}^{+1.68}$ TeV and $E_{c,p}=4.87\pm2.93$ TeV, for P1 and P2 respectively. When $E_{c,p}=10$ PeV is considered (blue dashed line in Fig. \ref{sed}), the data can be reproduced when $\alpha_{\rm p}=2.54\pm0.04$ and $\alpha_{\rm p}=2.46\pm0.09$ for P1 and P2, respectively. This modeling predicts emission also beyond the MAGIC data which can not be tested with the current instruments but certainly it is in agreement with the 100 MeV-400 GeV data. The cut-off energy $E_{c,p}=10$ PeV is an arbitrary value to show that when higher proton energies are considered the data can be reproduced, but cutoff at much higher energies cannot be excluded.\\
The modeling of the observed HE and VHE \gray{} data by $pp$ interactions allows to estimate the total energy of jet protons - perhaps a large fraction of the total jet power. The data can be reproduced when the total proton energy is $\int E_p N_{p}(E_p)dE_p\simeq(0.97-2.03)\times10^{53}$ erg corresponding to maximum luminosity of $L_{p} = W_{pp}/t_{pp}\simeq2.0\times10^{48}\:\rm{erg\:s^{-1}}$. In this case we note that for the cloud size of $5.2\times10^{13}$ cm and a number of density of $10^{10}\: {\rm cm^{-3}}$, the total energy of target protons is only a small fraction ($\sim10^{-5}$) of that of jet protons. $W_{pp}$ strongly depends on $n$ which is unknown and in this case is roughly constrained by the observed flux variation. However, even if it will differ from the used value by two orders of magnitude, the jet luminosity would be still realistic as it is measured in the rest frame. For example, the other hadronic models involving photo-meson interactions the required proton luminosity is much higher (e.g., $L_{p} \simeq10^{49}\:\rm{erg\:s^{-1}}$). The proton luminosity exceeds the power carried by the electrons $\sim 10^3$ times, making a significant contribution to the jet kinetic luminosity. Interestingly, this is similar to the difference between electron and proton luminosities usually estimated assuming one proton per electron. Clearly, both the predicted spectral shapes and the required total energy are physically reasonable, meaning that the $pp$ interactions can solely contribute to the observed HE and VHE \grays{}.\\
If the \grays{} observed from \bl{} are due to the interaction of protons in the target crossing the jet, secondary $e^{-}e^{+}$ pairs will be also produced whose emission can be significant in the X- or -\gray{} bands. As in the target the magnetic field most likely exceeds that in the jet, e.g., due to higher density of particles, these pairs dominantly will loose energy by synchrotron radiation (a sketch of a jet target interaction scenario can be seen in Fig. 1 of \cite{2012A&A...539A..69B}). In this case, the bremsstrahlung energy losses with a cooling time of $t_{\rm br}\simeq10^{16}/(ln(E_{e}/m_ec^2)+0.26)\:{\rm s}$ \citep{berezinskii} in a completely ionized medium is significantly longer than that of synchrotron emission (for $B\geq1$ G). The estimation of the magnetic field in the target is a rather difficult task, depending on the nature of the target, whether it is a cloud or a star envelope. This magnetic field is not necessarily constant, e.g., it can decrease due to expansion of the cloud or increase due to the continuous pumping of jet energy. Possible emission mechanisms when the cloud penetrates into the jet of M87 are discussed in \cite{2012ApJ...755..170B}. The blue dot-dashed line in Fig. \ref{sed} shows the synchrotron emission of secondary  $e^{-}e^{+}$ pairs in $\sim80$ G magnetic field. At lower energies, the boosted synchrotron emission of electrons directly accelerated in the jet is dominating but due to continuous electron energy losses by synchrotron and SSC radiations, this emission quickly drops already in the X-ray band. Being more energetic, the synchrotron emission from fresh $e^{-}e^{+}$ pairs can dominate beyond the X-ray band, reaching HEs. The synchrotron emission of $e^{-}e^{+}$ pairs {\it i)} can relatively well explain the X-ray spectrum observed during P1 when $B=80$ G (left panel in Fig. \ref{sed}) and {\it ii)} for the magnetic field $\sim800$ times stronger than that in the jet, it does not overproduce the data detected by Swift XRT during P2 (right panel in Fig. \ref{sed}) which are most likely due to synchrotron emission of jet electrons. It is interesting that the proton energy required to explain the observed HE and VHE \gray{} data is also sufficient for transferring enough energy to secondary electrons, so their synchrotron emission can explain the observed X-ray data for a reasonable magnetic field. This strengthens the used hadronic model. However, this X-ray emission in principle might arise also from the synchrotron emission of the secondary pairs produced in the cascades initiated by ultrahigh-energy protons \citep{magic}.\\
Having calculated the luminosity of protons and their energy distribution, the spectra of HE neutrinos can be calculated straightforwardly and a limit on the expected number of events detectable by IceCube from \bl{} in a certain exposure time ($t_{\rm exp}$) can be obtained. The neutrinos produced from the protons with an energy distribution of Eq. \ref{fin} will have a spectrum of $\sim E_{\nu}^{-\alpha_{\nu}}\exp(-\sqrt{ E_{\nu}/ E_{\nu,c}})$ where $\alpha_{\nu}\simeq\alpha_{p}-0.1$ and $E_{\nu,c}\simeq E_{c,p}/40$ \citep{kappes}. 
Therefore, even when $E_{c,p}=10$ PeV in the spectrum of neutrinos the cutoff is at $E_{\nu,c}\simeq250 \:{\rm TeV}$, close to the energy of the observed IceCube-170922A event. Next, using the effective area ($A_{\rm eff}(E_{\nu})$) most sensitive for the location of \bl{} from \citet{artsen17}, the number of expected events can be estimated as $N_{\nu}\simeq t_{\rm exp} \int A_{\rm eff}(E_{\nu})\Phi_{\nu}(E_{\nu})dE_{\nu}$. This effective area reaches its optimal (maximum) value for neutrino energies above several hundreds of TeV, so to estimate the expected neutrino rates, a significant impact will have the interactions of protons with $E_{\rm p}\geq1$ PeV. The number of expected events is proportional to the duration of the active emission phase of the source. The \gray{} light curve of \bl{} calculated above 2 GeV to avoid any bias from the nearby PKS 0502+049 is shown in Fig. 5 of \citep{sah}. As one can see the source was in its active phase around the IceCube-170922A event at least for $t_{\rm exp}>200$ days. Indeed, the \gray{} observations suggest that the active period can be $\sim 0.5-1$ year \citep{magic, IceCubeFermi, 2018arXiv180704537K}. Here we use $t_{\rm exp}=60$ days corresponding to the period when VHE \grays{} from \bl{} were observed and $t_{\rm exp}=0.5$ -year corresponding to the most prolonged active/bright state of the source. When $E_{c,p}=10$ PeV, the expected rates during P1 and P2 are $\simeq0.04$ and $\simeq0.15$ respectively for $t_{\rm exp}=60$ days, and $\simeq0.13$ and $\simeq0.46$ respectively for $t_{\rm exp}=0.5$ -year. Yet, assuming the proton acceleration continues beyond $10$ PeV or the source active phase is longer, the expected rate would be even higher.
\section{Conclusions}\label{sec:5}
In blazar studies, one of the long-standing and unclear questions is whether the protons are effectively accelerated in their jets and if they have a significant contribution to the observed emissions. The \gray{} observations solely are not sufficient to differentiate between the emission from electrons and protons not allowing to estimate their content in jets exactly. The recent association of \bl{} with the neutrino events allowed to measure the total energy of the jet carried by electrons and protons.\\
When the observed \grays{} and neutrinos from a blazar are due to $pp$ interactions, the energy of protons is mostly released in the GeV band allowing straightforward measurement of the proton spectra based on the observed \gray{} data. A simplified scenario of lepto-hadronic emission from \bl{} is discussed assuming that beside the constant boosted electron synchrotron/SSC emission from the jet compact region, a significant radiation in the \gray{} band is produced when a target (cloud, star envelope, etc.) crosses the jet and the inelastic $pp$ interactions produce pions which then decay into \grays{}. If only the emission from the leptons is considered, the electrons and magnetic field are in equipartition and the observed low-energy and time-averaged \gray{} data can be explained for the jet luminosity of $L_{\rm jet}\simeq 10^{45}\:{\rm erg\:s^{-1}}$. The \gray{} data when the neutrino was observed, can be modeled if the protons are distributed as $\sim E_{p}^{-2.50}$ and their energy extends up to $E_{c,p}=10$ PeV. The expected neutrino rate is $\simeq0.13-0.46$ during the long active phase of the source and $\sim 0.04-0.15$ events if the activity lasts 60 days. The synchrotron emission of electrons directly accelerated in the jet is significant up to the X-ray band, whereas the synchrotron emission of newly injected fresh pairs (from $pp$ interactions) in a dense target dominates afterwards explaining the observed X-ray data obtained during the low VHE \gray{} emission state of \bl{}. Within this scenario, the energy content of the protons (above $>$ GeV) in the blazar jet is estimated for the first time: the required proton injection luminosity should be $\simeq2.0\times10^{48}\:{\rm erg\:s^{-1}}$ which exceeds $10^{3}$ times that of electrons. This implies that a significant fraction of the jet kinetic energy is carried by the protons but still involvement of hadrons acceleration in the jet will not dramatically (unreasonably) increase its luminosity.  Considering the applied model can satisfactorily reproduce the observed multiwavelegth emission spectrum of \bl{} and predicts a sufficient neutrino production rate, it provides an acceptable explanation for the hadronic emission from the \bl{} jet.
\section*{acknowledgements}
We thank the anonymous referee for constructive comments.
\bibliography{references}

\begin{thebibliography}{}
\expandafter\ifx\csname natexlab\endcsname\relax\def\natexlab#1{#1}\fi

\bibitem[{{Aartsen} {et~al.}(2017{\natexlab{a}}){Aartsen}, {Abraham},
  {Ackermann}, \& et~al.}]{artsen17}
{Aartsen}, M.~G., {Abraham}, K., {Ackermann}, M., \& et~al. 2017{\natexlab{a}},
  \apj, 835, 151

\bibitem[{{Aartsen} {et~al.}(2017{\natexlab{b}}){Aartsen}, {Abraham},
  {Ackermann}, \& et~al.}]{aartsenB}
---. 2017{\natexlab{b}}, \apj, 835, 45

\bibitem[{{Aartsen} {et~al.}(2013){Aartsen}, {Abbasi}, {Abdou}, {Ackermann},
  {Adams}, {Aguilar}, {Ahlers}, {Altmann}, {Auffenberg}, {Bai}, \&
  et~al.}]{neutrino1}
{Aartsen}, M.~G., {Abbasi}, R., {Abdou}, Y., {et~al.} 2013, Physical Review
  Letters, 111, 021103

\bibitem[{{Aartsen} {et~al.}(2014){Aartsen}, {Ackermann}, {Adams}, {Aguilar},
  {Ahlers}, {Ahrens}, {Altmann}, {Anderson}, {Arguelles}, {Arlen}, \&
  et~al.}]{neutrino3}
{Aartsen}, M.~G., {Ackermann}, M., {Adams}, J., {et~al.} 2014, Physical Review
  Letters, 113, 101101

\bibitem[{{Aharonian}(2000)}]{ahartev}
{Aharonian}, F.~A. 2000, \na, 5, 377

\bibitem[{{Aharonian} {et~al.}(2017){Aharonian}, {Barkov}, \&
  {Khangulyan}}]{2017ApJ...841...61A}
{Aharonian}, F.~A., {Barkov}, M.~V., \& {Khangulyan}, D. 2017, \apj, 841, 61

\bibitem[{{Aharonian} {et~al.}(2002){Aharonian}, {Belyanin}, {Derishev},
  {Kocharovsky}, \& {Kocharovsky}}]{2002PhRvD..66b3005A}
{Aharonian}, F.~A., {Belyanin}, A.~A., {Derishev}, E.~V., {Kocharovsky}, V.~V.,
  \& {Kocharovsky}, V.~V. 2002, \prd, 66, 023005

\bibitem[{{Ahnen} {et~al.}(2018){Ahnen}, {Ansoldi}, {Antonelli}, {Arcaro},
  {Baack}, {Babi{\'c}}, {Banerjee}, {Bangale}, {Barres de Almeida}, {Abel
  Barrio}, {Becerra Gonz{\'a}lez}, {Bednarek}, {Bernardini}, {Berti},
  {Bhattacharyya}, {Biland}, {Blanch}, {Bonnoli}, {Carosi}, {Carosi},
  {Ceribella}, {Chatterjee}, {Merve Colak}, {Colin}, {Colombo}, {Contreras},
  {Cortina}, {Covino}, {Cumani}, {Da Vela}, {Dazzi}, {De Angelis}, {De Lotto},
  {Delfino}, {Delgado}, {Di Pierro}, {Dom{\'{\i}}nguez}, {Dominis Prester},
  {Dorner}, {Doro}, {Einecke}, {Elsaesser}, {Fallah Ramazani},
  {Fern{\'a}ndez-Barral}, {Fidalgo}, {Foffano}, {Pfrang}, {Fonseca}, {Font},
  {Fruck}, {Galindo}, {Gallozzi}, {Garc{\'{\i}}a L{\'o}pez}, {Garczarczyk},
  {Gaug}, {Giammaria}, {Godinovi{\'c}}, {Gora}, {Guberman}, {Hadasch}, {Hahn},
  {Hassan}, {Hayashida}, {Herrera}, {Hose}, {Hrupec}, {Inoue}, {Ishio},
  {Iwamura}, {Konno}, {Kubo}, {Kushida}, {Lelas}, {Lindfors}, {Lombardi},
  {Longo}, {L{\'o}pez}, {Maggio}, {Majumdar}, {Makariev}, {Maneva},
  {Manganaro}, {Mannheim}, {Maraschi}, {Mariotti}, {Mart{\'{\i}}nez}, {Masuda},
  {Mazin}, {Minev}, {Miranda}, {Mirzoyan}, {Moralejo}, {Moreno}, {Moretti},
  {Nagayoshi}, {Neustroev}, {Niedzwiecki}, {Nievas Rosillo}, {Nigro},
  {Nilsson}, {Ninci}, {Nishijima}, {Noda}, {Nogu{\'e}s}, {Paiano}, {Palacio},
  {Paneque}, {Paoletti}, {Paredes}, {Pedaletti}, {Peresano}, {Persic}, {Prada
  Moroni}, {Prandini}, {Puljak}, {Rodriguez Garcia}, {Reichardt}, {Rhode},
  {Rib{\'o}}, {Rico}, {Righi}, {Rugliancich}, {Saito}, {Satalecka},
  {Schweizer}, {Sitarek}, {Snidari{\'c}}, {Sobczynska}, {Stamerra}, {Strzys},
  {Suri{\'c}}, {Takahashi}, {Tavecchio}, {Temnikov}, {Terzi{\'c}}, {Teshima},
  {Torres-Alb{\`a}}, {Treves}, {Tsujimoto}, {Vanzo}, {Vazquez Acosta}, {Vovk},
  {Ward}, {Will}, {Zari{\'c}}, \& {Cerruti}}]{magic}
{Ahnen}, M.~L., {Ansoldi}, S., {Antonelli}, L.~A., {et~al.} 2018, ArXiv
  e-prints, arXiv:1807.04300

\bibitem[{{Araudo} {et~al.}(2010){Araudo}, {Bosch-Ramon}, \&
  {Romero}}]{araudo10}
{Araudo}, A.~T., {Bosch-Ramon}, V., \& {Romero}, G.~E. 2010, \aap, 522, A97

\bibitem[{{Araudo} {et~al.}(2013){Araudo}, {Bosch-Ramon}, \&
  {Romero}}]{araudo13}
---. 2013, \mnras, 436, 3626

\bibitem[{{Barkov} {et~al.}(2012{\natexlab{a}}){Barkov}, {Aharonian},
  {Bogovalov}, {Kelner}, \& {Khangulyan}}]{2012ApJ...749..119B}
{Barkov}, M.~V., {Aharonian}, F.~A., {Bogovalov}, S.~V., {Kelner}, S.~R., \&
  {Khangulyan}, D. 2012{\natexlab{a}}, \apj, 749, 119

\bibitem[{{Barkov} {et~al.}(2010){Barkov}, {Aharonian}, \&
  {Bosch-Ramon}}]{barkov}
{Barkov}, M.~V., {Aharonian}, F.~A., \& {Bosch-Ramon}, V. 2010, \apj, 724, 1517

\bibitem[{{Barkov} {et~al.}(2012{\natexlab{b}}){Barkov}, {Bosch-Ramon}, \&
  {Aharonian}}]{2012ApJ...755..170B}
{Barkov}, M.~V., {Bosch-Ramon}, V., \& {Aharonian}, F.~A. 2012{\natexlab{b}},
  \apj, 755, 170

\bibitem[{{Beall} \& {Bednarek}(1999)}]{beall}
{Beall}, J.~H., \& {Bednarek}, W. 1999, \apj, 510, 188

\bibitem[{{Bednarek}(2016)}]{bednarekneutrino}
{Bednarek}, W. 2016, \apj, 833, 279

\bibitem[{{Bednarek} \& {Banasi{\'n}ski}(2015)}]{bednarek15}
{Bednarek}, W., \& {Banasi{\'n}ski}, P. 2015, \apj, 807, 168

\bibitem[{{Bednarek} \& {Protheroe}(1997)}]{bednarek97}
{Bednarek}, W., \& {Protheroe}, R.~J. 1997, \mnras, 287, L9

\bibitem[{{Berezinskii} {et~al.}(1990){Berezinskii}, {Bulanov}, {Dogiel}, \&
  {Ptuskin}}]{berezinskii}
{Berezinskii}, V.~S., {Bulanov}, S.~V., {Dogiel}, V.~A., \& {Ptuskin}, V.~S.
  1990, {Astrophysics of cosmic rays}

\bibitem[{{B{\l}a{\.z}ejowski} {et~al.}(2000){B{\l}a{\.z}ejowski}, {Sikora},
  {Moderski}, \& {Madejski}}]{blazejowski}
{B{\l}a{\.z}ejowski}, M., {Sikora}, M., {Moderski}, R., \& {Madejski}, G.~M.
  2000, \apj, 545, 107

\bibitem[{{Bloom} \& {Marscher}(1996)}]{bloom}
{Bloom}, S.~D., \& {Marscher}, A.~P. 1996, \apj, 461, 657

\bibitem[{{Bosch-Ramon} {et~al.}(2012){Bosch-Ramon}, {Perucho}, \&
  {Barkov}}]{2012A&A...539A..69B}
{Bosch-Ramon}, V., {Perucho}, M., \& {Barkov}, M.~V. 2012, \aap, 539, A69

\bibitem[{{B{\"o}ttcher} \& {Dermer}(2010)}]{dermer10}
{B{\"o}ttcher}, M., \& {Dermer}, C.~D. 2010, \apj, 711, 445

\bibitem[{{Dar} \& {Laor}(1997)}]{dar}
{Dar}, A., \& {Laor}, A. 1997, \apjl, 478, L5

\bibitem[{{de la Cita} {et~al.}(2016){de la Cita}, {Bosch-Ramon},
  {Paredes-Fortuny}, {Khangulyan}, \& {Perucho}}]{cita}
{de la Cita}, V.~M., {Bosch-Ramon}, V., {Paredes-Fortuny}, X., {Khangulyan},
  D., \& {Perucho}, M. 2016, \aap, 591, A15

\bibitem[{{Dom{\'{\i}}nguez} {et~al.}(2011){Dom{\'{\i}}nguez}, {Primack},
  {Rosario}, {Prada}, {Gilmore}, {Faber}, {Koo}, {Somerville},
  {P{\'e}rez-Torres}, {P{\'e}rez-Gonz{\'a}lez}, {Huang}, {Davis},
  {Guhathakurta}, {Barmby}, {Conselice}, {Lozano}, {Newman}, \&
  {Cooper}}]{dominguez}
{Dom{\'{\i}}nguez}, A., {Primack}, J.~R., {Rosario}, D.~J., {et~al.} 2011,
  \mnras, 410, 2556

\bibitem[{{Georganopoulos} \& {Kazanas}(2003)}]{georganopoulos}
{Georganopoulos}, M., \& {Kazanas}, D. 2003, \apjl, 594, L27

\bibitem[{{Ghisellini} {et~al.}(1985){Ghisellini}, {Maraschi}, \&
  {Treves}}]{ghisellini}
{Ghisellini}, G., {Maraschi}, L., \& {Treves}, A. 1985, \aap, 146, 204

\bibitem[{{Ghisellini} \& {Tavecchio}(2009)}]{ghiselini09}
{Ghisellini}, G., \& {Tavecchio}, F. 2009, \mnras, 397, 985

\bibitem[{{Ghisellini} \& {Tavecchio}(2015)}]{ghistav}
---. 2015, \mnras, 448, 1060

\bibitem[{{Giannios} {et~al.}(2009){Giannios}, {Uzdensky}, \&
  {Begelman}}]{giannios09}
{Giannios}, D., {Uzdensky}, D.~A., \& {Begelman}, M.~C. 2009, \mnras, 395, L29

\bibitem[{{Giannios} {et~al.}(2010){Giannios}, {Uzdensky}, \&
  {Begelman}}]{giannios10}
---. 2010, \mnras, 402, 1649

\bibitem[{{IceCube} {et~al.}(2018){IceCube}, {Fermi-LAT}, {MAGIC}, {AGILE},
  {ASAS-SN}, {HAWC}, {S}, {INTEGRAL}, {Kanata}, {Kiso}, {Kapteyn}, {telescope},
  {Subaru}, {Swift/NuSTAR}, {VERITAS}, \& {VLA/17B-403 teams}}]{IceCubeFermi}
{IceCube}, T., {Fermi-LAT}, {MAGIC}, {et~al.} 2018, ArXiv e-prints,
  arXiv:1807.08816

\bibitem[{{IceCube Collaboration}(2013)}]{neutrino2}
{IceCube Collaboration}. 2013, Science, 342, 1242856

\bibitem[{{IceCube Collaboration}(2018)}]{IceCube1}
---. 2018, ArXiv e-prints, arXiv:1807.08794

\bibitem[{{Inoue} \& {Takahara}(1996)}]{inou}
{Inoue}, S., \& {Takahara}, F. 1996, \apj, 463, 555

\bibitem[{{Joshi} \& {B{\"o}ttcher}(2011)}]{joshi}
{Joshi}, M., \& {B{\"o}ttcher}, M. 2011, \apj, 727, 21

\bibitem[{{Kadler} {et~al.}(2016){Kadler}, {Krau{\ss}}, {Mannheim}, \&
  et~al.}]{kadler}
{Kadler}, M., {Krau{\ss}}, F., {Mannheim}, K., \& et~al. 2016, Nature Physics,
  12, 807

\bibitem[{{Kappes} {et~al.}(2007){Kappes}, {Hinton}, {Stegmann}, \&
  {Aharonian}}]{kappes}
{Kappes}, A., {Hinton}, J., {Stegmann}, C., \& {Aharonian}, F.~A. 2007, \apj,
  656, 870

\bibitem[{{Keivani} {et~al.}(2018){Keivani}, {Murase}, {Petropoulou}, {Fox},
  {Cenko}, {Chaty}, {Coleiro}, {DeLaunay}, {Dimitrakoudis}, {Evans}, {Kennea},
  {Marshall}, {Mastichiadis}, {Osborne}, {Santander}, {Tohuvavohu}, \&
  {Turley}}]{2018arXiv180704537K}
{Keivani}, A., {Murase}, K., {Petropoulou}, M., {et~al.} 2018, ArXiv e-prints,
  arXiv:1807.04537

\bibitem[{{Kelner} {et~al.}(2006){Kelner}, {Aharonian}, \&
  {Bugayov}}]{kelner_06}
{Kelner}, S.~R., {Aharonian}, F.~A., \& {Bugayov}, V.~V. 2006, \prd, 74, 034018

\bibitem[{{Khangulyan} {et~al.}(2013){Khangulyan}, {Barkov}, {Bosch-Ramon},
  {Aharonian}, \& {Dorodnitsyn}}]{2013ApJ...774..113K}
{Khangulyan}, D.~V., {Barkov}, M.~V., {Bosch-Ramon}, V., {Aharonian}, F.~A., \&
  {Dorodnitsyn}, A.~V. 2013, \apj, 774, 113

\bibitem[{{Khiali} \& {de Gouveia Dal Pino}(2016)}]{2016MNRAS.455..838K}
{Khiali}, B., \& {de Gouveia Dal Pino}, E.~M. 2016, \mnras, 455, 838

\bibitem[{{Mannheim}(1995)}]{man}
{Mannheim}, K. 1995, Astroparticle Physics, 3, 295

\bibitem[{{Mannheim} \& {Biermann}(1992)}]{mannheim}
{Mannheim}, K., \& {Biermann}, P.~L. 1992, \aap, 253, L21

\bibitem[{{Maraschi} {et~al.}(1992){Maraschi}, {Ghisellini}, \&
  {Celotti}}]{maraschi}
{Maraschi}, L., {Ghisellini}, G., \& {Celotti}, A. 1992, \apjl, 397, L5

\bibitem[{{Marscher} \& {Gear}(1985)}]{marscher}
{Marscher}, A.~P., \& {Gear}, W.~K. 1985, \apj, 298, 114

\bibitem[{{M{\"u}cke} \& {Protheroe}(2001)}]{muckesynch}
{M{\"u}cke}, A., \& {Protheroe}, R.~J. 2001, Astroparticle Physics, 15, 121

\bibitem[{{Murase} {et~al.}(2016){Murase}, {Guetta}, \&
  {Ahlers}}]{2016PhRvL.116g1101M}
{Murase}, K., {Guetta}, D., \& {Ahlers}, M. 2016, Physical Review Letters, 116,
  071101

\bibitem[{{Netzer}(2015)}]{2015ARA&A..53..365N}
{Netzer}, H. 2015, \araa, 53, 365

\bibitem[{{Padovani} {et~al.}(2018){Padovani}, {Giommi}, {Resconi}, {Glauch},
  {Arsioli}, {Sahakyan}, \& {Huber}}]{sah}
{Padovani}, P., {Giommi}, P., {Resconi}, E., {et~al.} 2018, ArXiv e-prints,
  arXiv:1807.04461

\bibitem[{Padovani \& Resconi(2014)}]{padovani1}
Padovani, P., \& Resconi, E. 2014, Monthly Notices of the Royal Astronomical
  Society, 443, 474

\bibitem[{Padovani {et~al.}(2016)Padovani, Resconi, Giommi, Arsioli, \&
  Chang}]{padovani2}
Padovani, P., Resconi, E., Giommi, P., Arsioli, B., \& Chang, Y.~L. 2016,
  Monthly Notices of the Royal Astronomical Society, 457, 3582

\bibitem[{{Paiano} {et~al.}(2018){Paiano}, {Falomo}, {Treves}, \&
  {Scarpa}}]{paiano}
{Paiano}, S., {Falomo}, R., {Treves}, A., \& {Scarpa}, R. 2018, \apjl, 854, L32

\bibitem[{{Perucho} {et~al.}(2017){Perucho}, {Bosch-Ramon}, \&
  {Barkov}}]{2017A&A...606A..40P}
{Perucho}, M., {Bosch-Ramon}, V., \& {Barkov}, M.~V. 2017, \aap, 606, A40

\bibitem[{{Peterson}(2006)}]{2006LNP...693...77P}
{Peterson}, B.~M. 2006, in Lecture Notes in Physics, Berlin Springer Verlag,
  Vol. 693, Physics of Active Galactic Nuclei at all Scales, ed. D.~{Alloin},
  77

\bibitem[{{Protheroe}(1997)}]{prot}
{Protheroe}, R.~J. 1997, in Astronomical Society of the Pacific Conference
  Series, Vol. 121, IAU Colloq. 163: Accretion Phenomena and Related Outflows,
  ed. D.~T. {Wickramasinghe}, G.~V. {Bicknell}, \& L.~{Ferrario}, 585

\bibitem[{{Sahakyan} {et~al.}(2014){Sahakyan}, {Piano}, \& {Tavani}}]{sahakyan}
{Sahakyan}, N., {Piano}, G., \& {Tavani}, M. 2014, \apj, 780, 29

\bibitem[{{Sikora} {et~al.}(1994){Sikora}, {Begelman}, \& {Rees}}]{sikora}
{Sikora}, M., {Begelman}, M.~C., \& {Rees}, M.~J. 1994, \apj, 421, 153

\bibitem[{{Sikora} {et~al.}(2009){Sikora}, {Stawarz}, {Moderski}, {Nalewajko},
  \& {Madejski}}]{sikora09}
{Sikora}, M., {Stawarz}, {\L}., {Moderski}, R., {Nalewajko}, K., \& {Madejski},
  G.~M. 2009, \apj, 704, 38

\bibitem[{{Sokolov} {et~al.}(2004){Sokolov}, {Marscher}, \&
  {McHardy}}]{sokolov}
{Sokolov}, A., {Marscher}, A.~P., \& {McHardy}, I.~M. 2004, \apj, 613, 725

\bibitem[{{Spada} {et~al.}(2001){Spada}, {Ghisellini}, {Lazzati}, \&
  {Celotti}}]{spada}
{Spada}, M., {Ghisellini}, G., {Lazzati}, D., \& {Celotti}, A. 2001, \mnras,
  325, 1559

\bibitem[{{Tavecchio} \& {Ghisellini}(2008)}]{tavecchio08}
{Tavecchio}, F., \& {Ghisellini}, G. 2008, \mnras, 385, L98

\bibitem[{{Tavecchio} \& {Ghisellini}(2015)}]{tavec}
---. 2015, \mnras, 451, 1502

\bibitem[{{Urry} \& {Padovani}(1995)}]{urry}
{Urry}, C.~M., \& {Padovani}, P. 1995, \pasp, 107, 803

\bibitem[{{Wang} \& {Liu}(2016)}]{2016PhRvD..93h3005W}
{Wang}, X.-Y., \& {Liu}, R.-Y. 2016, \prd, 93, 083005

\bibitem[{{Yoshida} {et~al.}(1997){Yoshida}, {Dai}, {Jui}, \&
  {Sommers}}]{yoshida}
{Yoshida}, S., {Dai}, H., {Jui}, C.~C.~H., \& {Sommers}, P. 1997, \apj, 479,
  547

\bibitem[{{Zabalza}(2015)}]{zabalza}
{Zabalza}, V. 2015, in International Cosmic Ray Conference, Vol.~34, 34th
  International Cosmic Ray Conference (ICRC2015), 922

\end{thebibliography}
\bibliographystyle{aasjournal}

\end{document}